\documentclass[twocolumn,amsmath,amssymb,showpacs,prl,superscriptaddress]{revtex4-1}

\usepackage{epsfig}
\usepackage{color}
\usepackage{graphicx}
\usepackage{bm}
\usepackage{epstopdf}

\begin{document}

\newcommand{\bra}[1]{\langle #1 |}
\newcommand{\ket}[1]{| #1 \rangle}
\newcommand{\braket}[2]{\langle #1 | #2 \rangle}
\newcommand{\expv}[1]{\langle #1 \rangle}
\newcommand{\us}{\uparrow}
\newcommand{\ds}{\downarrow}
\newcommand{\hlf}{\frac{1}{2}}
\newcommand{\veps}{\varepsilon}
\newcommand{\bs}[1]{\boldsymbol{#1}}

\title{Simulation of Time-dependent Heisenberg Models in 1D}
 
\author{A.~G. \surname{Volosniev}}
\affiliation{Department of Physics and Astronomy, Aarhus University, DK-8000 Aarhus C,  Denmark}
\affiliation{Institut f{\"u}r Kernphysik, Technische Universit{\"a}t Darmstadt, 64289 Darmstadt, Germany}
\author{H.-W. \surname{Hammer}}
\affiliation{Institut f{\"u}r Kernphysik, Technische Universit{\"a}t Darmstadt, 64289 Darmstadt, Germany}
\affiliation{
ExtreMe Matter Institute EMMI, GSI Helmholtzzentrum f{\"u}r Schwerionenforschung GmbH, 64291 Darmstadt, Germany}
\author{N.~T. \surname{Zinner}}
\affiliation{Department of Physics and Astronomy, Aarhus University, DK-8000 Aarhus C,  Denmark}


\date{\today}

\begin{abstract}

In this paper, we provide a theoretical analysis of strongly interacting quantum systems confined by a time-dependent 
external potential in one spatial dimension. We show that such systems can be used to simulate spin chains described
by Heisenberg Hamiltonians in which the exchange coupling constants can be manipulated by time-dependent driving
of the shape of the external confinement.
As illustrative examples, we consider a harmonic trapping potential with a variable frequency and an infinite 
square well potential with a time-dependent barrier in the middle.
\end{abstract}

\maketitle

\noindent
{\it Introduction}. --
Strongly interacting quantum systems are an intricate and exciting part of theoretical physics. 
Their intricacy is due to the strong many-body correlations that 
may lead to unexpected new phenomena not present in the weakly-interacting case. 

For systems with strong interparticle coupling, one spatial dimension (1D) plays a very special 
role \cite{rigol2011}. One reason for this is the unusual duality, often called the Fermi-Bose mapping, 
between 1D impenetrable bosons and ideal fermions, which was rigorously shown in 1960 by 
Girardeau~\cite{girardeau1960}. The most exciting aspect of this duality is the possibility to study it in
modern experimental setups with two different atomic 
species~\cite{tg2004,Kinoshita20082004,jochim2012}. As a future perspective, the Fermi-Bose mapping 
suggests~\cite{volosniev2014,deuret2014,volosniev2015,levinsen2014} to engineer a chain of spins with 
adjustable nearest-neighbor couplings using a strongly repulsive multicomponent system in a 
trap~\cite{serwane2011,pagano2014,jochim2014}.
Such spin chains possess a very high degree of tunability thus opening the possibility of realizing and studying 
phenomena such as 1D $SU(N)$ quantum magnets and perfect state transfer~\cite{bose2003,volosniev2015}. 

While the Fermi-Bose mapping was first established for a stationary system, the generalization to the case of a time-dependent 
trapping potential is straighforward for a system of impenetrable bosons~\cite{wright2000,girardeau2000}.  To the best of our 
knowledge, however, such a generalization for multicomponent systems with large but finite interaction was 
not previously discussed in the literature \footnote{Recent studies have considered time-dependent driving starting from the lattice approximation \cite{itin2014}. However we note that our formalism does not require one to make a lattice model approximation.}. 
Due to the interplay of two different time-dependent effects, this generalization is far from obvious. First, there is the motion of particles due to the time-dependent trapping potential, and second, there is the particle exchange. As we will show the timescales for these effects are effectively decoupled from one another and the dynamics of particle exchange is determined by the trapping potential. 

\begin{figure}
\centerline{\includegraphics[scale=0.55]{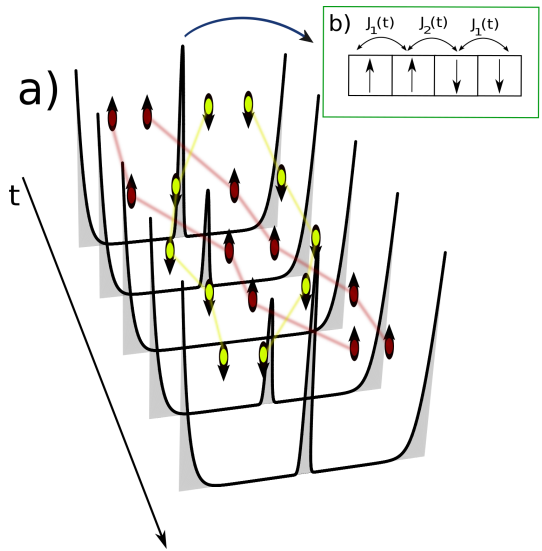}}
\caption{(Color Online). Mapping of a strongly interacting one-dimensional system in a time-dependent potential 
, a), onto a spin chain with time dependent coefficients, b).}
\label{fig:gen}
\end{figure}

In this paper, we consider a system with two kinds of spinless fermions with strong interspecies repulsion.  
We first show that the behaviour of such a system can be described by the Heisenberg Hamiltonian 
with time-dependent exchange coupling coefficients.  These coefficients can be altered by manipulating the shape of the trapping potential as a whole. 
This contrasts our idea with an idea of realizing a time-dependent Heisenberg Hamiltonian on a lattice by addressing every site independently.
As we discuss below, our approach has very different strengths and limitations and thus ideally complements the standard lattice approach.
In particular, it allows to address any trapping potential and is not
limited to the lattice approximation.
For a four-atom system this is sketched in Fig.~\ref{fig:gen}. Part a) shows 
the initial configuration with two fermions in one well and two fermions of a different kind in the other well. 
We also sketch a possible evolution of this configuration in a time-dependent potential where the final state 
corresponds to the exchange of the two pairs.
This evolution can be described by mapping the system to a spin-chain model described by a Heisenberg Hamiltonian
where the coupling coefficients depend on time, see Fig. \ref{fig:gen}~b).  The mapping opens a way to engineer 
and simulate driven Heisenberg Hamiltonians with time-dependent coefficients where the time dependence gives an extra knob to tune the dynamics in the system \cite{Bruder2006, Galve2009}.

To illustrate our findings, we apply this mapping to a time-dependent harmonic potential and an
infinite square well potential with a time-dependent barrier in the middle. For the former case, we show that the 
coupling coefficients in the spin chain are simply multiplied with a position independent scale factor.
In the latter case, one can tune the middle coupling coefficient almost independently from the others. This allows one to achieve a controlled exchange of pairs, see Fig. \ref{fig:gen}~a). 

{\it Formulation}. --
For the sake of the argument let us start with a 1D system of $N$ spinless fermions of one kind (spin up) and one fermion of another 
kind (spin down). We assume that every particle has mass $m$ and is confined by the same time-dependent trapping potential 
$\varepsilon V(x/L,\varepsilon t/\hbar)$, where $\varepsilon=\hbar^2/(mL^2)$ and $L$ is some natural time-independent unit of length. 
For convenience, we assume that $m=\hbar=L=1$ from now on. 

The dynamics of a system with a spin-up fermion placed at $x$ and spin-down 
fermions at $y_1,...,y_N$ is described by the wave function
$\Psi(x,y_1,...,y_N,t)$, which  satisfies the Schr{\"o}dinger equation,
\begin{equation}
i\frac{\partial}{\partial t}\Psi=H\Psi, \qquad H=\sum_{i=1}^N h(y_i,t) + h(x,t),
\label{Eq:Ham}
\end{equation}
where $h(x,t)=-\frac{1}{2}\frac{\partial^2}{\partial x^2}+V(x,t)$ is the one-body Hamiltonian. The zero-range interaction enters 
through 1D Bethe-Peierls boundary conditions at the points where the particles meet 
(see e.g. Ref.~\cite{McGuire1965}):
\begin{equation}
\left(\frac{\partial \Psi}{\partial x}-\frac{\partial \Psi}{\partial y_i}\right)_{x=y_i^+}-
\left(\frac{\partial \Psi}{\partial x}-\frac{\partial \Psi}{\partial y_i}\right)_{x=y_i^-}=
2g\Psi(x=y_i),
\label{eq:boundcond}
\end{equation}
where $g$ is the interaction strength and the notation $x=y_i^{\pm}$ means that the derivative is taken at the point  
$x=y_i\pm\varepsilon$, with $\varepsilon>0$ and the limit $\varepsilon \to 0$ is taken afterwards. Below, we consider the dynamics of 
the system in the following scenario: the interaction is adiabatically tuned in a constant trapping potential $V(x,0)$ from zero to 
some value $g^f$ which is very large. This procedure initializes the state $\Psi^{(0)}(x,y_1,...,y_N)$. It is assumed that at later times 
the shape of the trapping potential depends on time and we look for the wave function satisfying Eq.~(\ref{Eq:Ham}) with the initial 
condition $\Psi(t=0)=\Psi^{(0)}$.

{\it Initial state}. --
Let us start by discussing the initial wave function $\Psi^{(0)}$. If $1/g^f=0$ then Eq.~(\ref{eq:boundcond}) dictates that the 
particles cannot exchange their relative positions and $\Psi^{(0)}$ should be described separately on each ordering of particles, 
e.g. $x<y_1<y_2<...<y_N$, on which the solution is obtained from the Fermi-Bose mapping~\cite{girardeau1960}:
\begin{align}
\Psi^{(0)}=\sum_{j=1}^{N+1}a_j^{(0)}\chi_j(x,y_1,...,y_N)\Phi^{(0)}(x,y_1,...,y_N),
\label{eq:wavefunc0}
\end{align}
where  function $\chi_j$ is non-zero only if it contains $j-1$ arguments $y_i$ that are smaller than $x$, and $\Phi^{(0)}$ is one of 
the eigenstates of the Hamiltonian for $N+1$ spinless fermions (for the illustrative examples below we use the ground state).  
First note that the states from Eq.~(\ref{eq:wavefunc0}) are $N+1$-fold degenerate even if the eigenspectrum of spinless fermions is 
non-degenerate.  Thus, to find $\Psi^0$, we should find the adiabatic eigenstates in $g$, that are characterized by $a_j^{(0)}$. 
This can be done perturbatively by minimizing the energy in the limit $g_f\to\infty$~\cite{volosniev2014, volosniev2014a, blume2015}. For large but finite 
interaction strengths, the wave function preserves the form given by Eq.~(\ref{eq:wavefunc0}) but acquires an additional contribution 
proportional to $1/g_f$.  Furthermore, the minimization of energy leads to the mapping of a system onto a spin chain. To establish such a mapping in the time-dependent case, where the energy is not a good quantum number, a new approach is necessary and this is what we provide in this paper.

{\it Time dynamics}. --
At $t>0$, the external potential depends on time and the time evolution is described by $\Psi(t)$. Let us first consider the system 
with infinite interaction, i.e. $1/g_f=0$. In this case, the wave function at each ordering should still be described with the wave function 
of spinless fermions~\cite{wright2000}, $\Phi(x,y_1,...,y_N,t)$. Moreover, the probability of each ordering cannot be changed since the 
particles do not exchange their position. So $\Psi(t)$ in this limit has the same form as in Eq.~(\ref{eq:wavefunc0}) with $\Phi(t)$ 
instead of $\Phi^{(0)}$.

Let us now assume that the interaction strength is large but finite. Apparently this means that the wave function at each ordering cannot 
be described exactly with $\Phi^{(0)}$, and we should to look for a solution in the form $\Psi=\phi+\frac{1}{g_f}f$ where the function $\phi$ reads
\begin{align}
\phi(x,y_1,...,y_N,t)=\sum_{j=1}^{N+1}a_j(t)\chi_j(x,y_1,...,y_N)\Phi.
\label{eq:wavefunc}
\end{align} 
Without any loss of generality, we assume that $\int f^* \chi_i \Phi \,\mathrm{d}x\,\mathrm{d}y_1...\mathrm{d}y_N =0$,  i.e. that the functions 
$f$ and $\Phi$ are orthogonal on each ordering of the coordinates.
Having in mind these conditions, we insert Eq.~(\ref{eq:wavefunc}) 
in the Schr{\"o}dinger equation. To proceed further, we insert the ansatz wave function into the Schr{\"o}dinger equation and project it on each ordering. Next using that $f$ is orthogonal to $\Phi$ at every moment of time together with the boundary conditions (\ref{eq:boundcond}), we eliminate the function $f$  (See Supplemental Material \cite{supp}). This procedure allows us to 
obtain a system of equations for the set of coefficients, $a_j(t)$,
\begin{equation}
i\frac{\mathrm{d} a_j}{\mathrm{d} t}=a_j(J_{j-1}+J_j)-a_{j-1}J_{j-1}-a_{j+1}J_j + O\left(\frac{1}{g_f^2}\right),
\label{eq:ai}
\end{equation}
where, assuming that $J_0=J_{N+1}=0$, the parameters $J_i$ are defined as follows
\begin{equation}
J_j(t)=-\frac{1}{g_f}\frac{\int_{y_1<y_2<...<y_N}\mathrm{d}y_1...\mathrm{d}y_N \left|\frac{\partial \Phi(t)}{\partial x}\right|^2_{x=y_{j}}}{\int_{x<y_1<y_2<...<y_N}\mathrm{d}x\mathrm{d}y_1...\mathrm{d}y_N  \Phi^2(t)}.
\end{equation}
After writing  Eq.~(\ref{eq:ai}) in matrix form, it becomes apparent that up to the order $1/g^2_f$ this equation also describes the dynamics 
of a spin chain with the Heisenberg Hamiltonian 
\begin{equation}
H_s =- \hlf \sum_{j=1}^{N} 
J_j(t) (\bs{\sigma}^j \bs{\sigma}^{j+1} - \mathbf{I} ),
\label{Eq:HamSpin}
\end{equation}
and the corresponding wave function is
\begin{equation}
|F\rangle = \sum_{j=1}^{N+1}a_j(t)|\us_1...\ds_j...\us_{N+1}\rangle,
\label{Eq:WaveSpin}
\end{equation}
where we denote the identity operator on every site with~$\mathbf{I}$, 
$\bs{\sigma}^j=(\sigma^j_x,\sigma^j_y,\sigma^j_z)$ are the Pauli 
matrices acting on a spin at site $j$, 
and $J_j$ are site- and time- dependent interaction coefficients.
Equations~(\ref{Eq:HamSpin}) and (\ref{Eq:WaveSpin}) generalize the time-independent mapping~\cite{volosniev2014, deuret2014, volosniev2015} 
onto a spin-chain Hamiltonian to the time-dependent case.   
 The derivation above implies that the time scale for the particle motion in leading order (in $1/g_f$) is determined by the trap alone, whereas the time scale of the spin exchange is proportional to $1/g_f$. It is related to the famous spin-charge separation \cite{giamarchi-book} in 1D, although here we derived it for a strongly interacting mesoscopic system from first principles in the presence of an external potential that depends on time. Note that there are  
higher order contributions to both, the particle motion and spin exchange. 
However, these corrections are negligible in the case of strong interactions,
$1/g_f\to 0$, and therefore we do not need to consider them here.

Applying the presented approach it is easy to show that the Hamiltonian (\ref{Eq:HamSpin}), 
can be used for any number of spin-down fermions similar to the time-independent case (see Ref.~\cite{volosniev2014} for a derivation).  
This is due to the fact that the main process in the system is 
the spin exchange of neighboring particles which is correctly described in the Hamiltonian (\ref{Eq:HamSpin}). 
The same logic also applies to multicomponent system or systems made of strongly-interacting bosons. 

{\it Discussion}. --
We first assume that the coupling coefficients, $J_i$, are independent of time.  Then 
linear system of equations (\ref{eq:ai}) has the fundemental set of solutions: $a_j(t)=a^{(0)}_je^{-i\epsilon t}$, where $\epsilon$ is the relevant 
eigenvalue of the Hamiltonian (\ref{Eq:HamSpin}). Let us now consider what happens if the external trapping potential depends on time. 
To find the coefficients $J_i$ in this case, we first need to solve a time-dependent one-body problem and construct a Slater determinant wave 
function $\Phi$ out of the established solutions. 

As our first application, we consider a system trapped by a harmonic oscillator  potential,  $V(x,t)=\omega^2(t)x^2/2$, \
for which one-body solutions are known~\cite{Husimi1953,perelomov1969,castin2004,moroz2012}, yielding $\Phi(x,y_1,...,y_N,t)$ from $\Phi^{(0)}$ as
\begin{equation}
\Phi=\frac{e^{-i E\int_0^t\frac{\mathrm{d}\tau}{\lambda^2(\tau)}}}{\sqrt{\lambda(t)^{N+1}}}e^{i\left(x^2+\sum_{i=1}^N y_i^2\right)\frac{\dot\lambda}{2\lambda}} \Phi^{(0)}\left(\frac{x}{\lambda},...,\frac{y_N}{\lambda}\right),
\end{equation}
where $E$ is the initial energy and
$\lambda(t)$ is the time-dependent scale parameter.
Its time derivative $\dot\lambda$ is determined from the equation:
$\lambda^3\ddot\lambda=1-\omega^2(t)\lambda^4(t)$.
Since our choice of units sets $\omega^2(t<0)=1$, the initial conditions for this equation read $\lambda(t<0)=1$ and 
$\dot\lambda(t<0)=0$. Obviously, such a wave function $\Phi$ produces $J_i(t)\sim1/\lambda^3(t)$,
so that all coupling constants depend on time in the same way. The corresponding system of equations (\ref{eq:ai}) has the 
following fundamental set of solutions:
$ a_j(t)=a_j^{(0)}e^{-i\epsilon\int_0^{t}\frac{\mathrm{d}\tau}{\lambda^3(\tau)}}$.
Thus we see that the scale invariance given by the harmonic trap is preserved up to terms suppressed by $1/g_f$ in the form:
\begin{equation}
\Psi=\frac{e^{-i\epsilon\int_0^{t}\frac{\mathrm{d}\tau}{\lambda^3(\tau)}-i E\int_0^t\frac{\mathrm{d}\tau}{\lambda^2(\tau)}}}{\sqrt{\lambda(t)^{N+1}}e^{-i\left(x^2+\sum_{i=1}^N y_i^2\right)\frac{\dot\lambda}{2\lambda}}} \Psi^{(0)}\left(\frac{x}{\lambda},...,\frac{y_N}{\lambda}\right).
\end{equation}
Therefore, the overall spin dynamics in the system is not affected 
by a change of the external potential up to corrections
suppressed by $1/g_f$.
Of course, the harmonic oscillator is a truly special case due to the scale invariance and any trapping potential 
that is not scale invariant will have more pronounced effects on the system. 
A detailed discussion of the breaking of scale invariance 
in the oscillator for two particles by higher order corrections can be found in 
Ref.~\cite{Ebert:2015gko}.

It is interesting to note that the spin dynamics of the system in a harmonic trap can be altered by a
time-independent weak magnetic field \cite{volosniev2015}.
With a magnetic field the Hamiltonian is 
$\tilde{H} = H + \sum_{i=1}^{N} \frac{b(y_i)}{g_f} - \frac{b(x)}{g_f}$, where, for simplicity, we again consider a system with only one spin-down
fermion \footnote{ Notice, that the resulting Hamiltonian can be straighforwardly generalized to more particles}. The correponding spin chain Hamiltonian is written as
\begin{equation}
\tilde{H}_s = - \hlf \sum_{j=1}^{N} 
J_j (\bs{\sigma}^j \bs{\sigma}^{j+1} - \mathbf{I} ) + \sum_{j=1}^{N+1} \beta_j \sigma_z^j,
\label{Eq:HamSpinFerm}
\end{equation}
where
\begin{equation}
\beta_j(t) = \frac{1}{g_f}\frac{\int \mathrm{d}y_1...\mathrm{d}y_N \mathrm{d} x \chi_j(x,y_1,...,y_N)
|\Phi|^2 \, b(x)} 
{\int \mathrm{d}y_1...\mathrm{d}y_N \mathrm{d} x \chi_j(x,y_1,...,y_N)
|\Phi|^2}.
\end{equation}
Notice that coefficients $\beta_j$ depend on time through $\Phi$ even though the  magnetic field is stationary. However, since this time dependence can be different from
$J_i(t)$ the probability of each ordering in the spin chain, i.e. $|a_i(t)|^2$ changes with time. An application of magnetic field can drive a transition to a spin 
segregated state where the spin down particles are not mixed with the spin up particles.
As was discussed in Ref.~\cite{cui2015}, this is possible due to a high degeneracy of the spectrum such that even a tiny magnetic field gradient can drive such a transition. 
This can be utilized using the magnetic field in the form $b(x)=b_0 x$, where $b_0$ is some constant parameter (therefore $\beta_j=\lambda(t)\beta_j^{(0)}$). By taking this constant to be  large, such that $\beta_j(0)/J_j(0)\gg1$, we can have the initial state to be almost fully spin segregated or 'ferromagnetic'. By increasing the frequency 
of the external confinement we can drive the system from dominantly 'ferromagnetic' to 'antiferromagnetic' states, since the 
Heisenberg Hamiltonian in Eq. (\ref{Eq:HamSpin}) is 'antiferromagnetic'.

\begin{figure}[t]
\centerline{\includegraphics[scale=0.7]{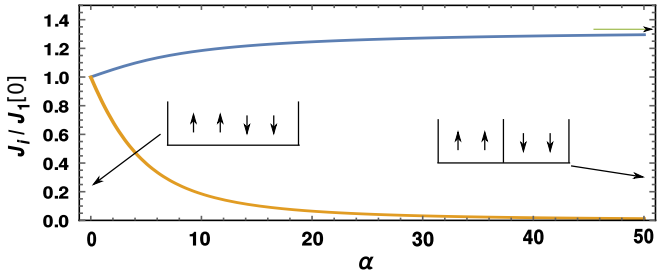}}
\caption{(Color Online). The coupling constants, $J_i/J_1(0)$ as functions of the barrier height, $\alpha$. The upper thick (blue) curve describes $J_1$, the lower (orange) curve corresponds to $J_2$. With the upper most arrow, we show the limiting value of $J_1$ for $\alpha\to\infty$. The insets show the system for the corresponding values of $\alpha$.}
\label{fig:ratio}
\end{figure}

To conclude the presentation of the formalism, we consider a trapping potential where the quantum dynamics of a spin chain is altered without 
applying an external magnetic field. 
For this we use a potential schematically shown in Fig.~\ref{fig:gen}, where a shallow area  with a time-dependent barrier in the center is surrounded 
by impenetrable wells. We model this trap by an infinite square well 
potential, i.e. $V(x,t)=\alpha f(t)\delta(x)$ for $x\in[-1,1]$ and otherwise $V(x,t)\to\infty$. 
To give a spin chain time to react on the change of potential,  we assume that $f(t)$ varies significantly only on a time scale given by $g_f$. 
This assumption means that $\Phi(t)$ changes almost adiabatically, which however does not imply adiabatic change of $\Psi(t)$ 
due to the degeneracy of the spectrum. Having this in mind, let us first assume that $f(t)=1$ and study $J_i$ for different $\alpha$. Note that 
conservation of parity leads to $J_1[\alpha]=J_3[\alpha]$. So it is enough to study only the combinations 
$J_1[\alpha]/J_1[0]$ and $J_2[\alpha]/J_1[0]$
which are $g_f$-independent and are shown in Fig.~\ref{fig:ratio}.  For $\alpha=0$, we have a pure infinite square well potential which requires $J_1=J_2$.
Positive values of $\alpha$ naturally descrease $J_2$ and increase $J_1$, such that for $\alpha \to \infty$, 
we have $J_2/J_1[0]\to 0$ and $J_1/J_1[0]\to4/3$. The increase of $J_1[\alpha]/J_1[0]$ is related to the increase of the density in one well by 
increasing the barrier. Note that this effect should be less visible for more particles.

\begin{figure}[t]
\centerline{\includegraphics[scale=0.5]{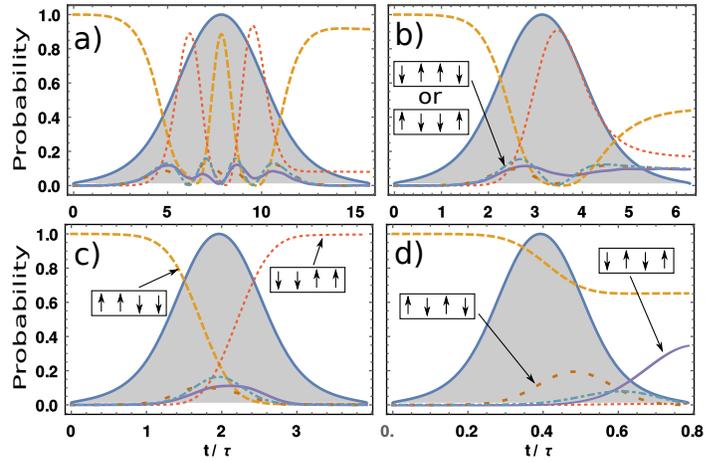}}
\caption{(Color Online.) Dynamics of probabilities for different configurations as a function of time for a four body system in an infinite square well potential with a time-independent barrier in the middle. Panel a) corresponds to $A=0.2$, b) \ $A=0.5$, c) \ $A=0.8$ and d) \ $A=4$. The solid (blue) shaded curve shows the change of $J_2/J_1[0]$ as a function of time. Dashed (orange) curve depicts $\uparrow\uparrow\downarrow\downarrow$ configuration, dotted (red) - $\downarrow\downarrow\uparrow\uparrow$. These and all other configurations are  shown in panels b), c) and d). Note that $\uparrow\downarrow\downarrow\uparrow$ and $\downarrow\uparrow\uparrow\downarrow$ configurations are equally probable.}
\label{fig:ferr}
\end{figure}

To illustrate the effect of this change of $J_i[\alpha]/J_1[0]$, we assume that for $\alpha\to\infty$ we prepare the system in the $\uparrow\uparrow\downarrow\downarrow$ 
configuration, see Fig. \ref{fig:gen}. Now we open and close the barrier and investigate the evolution of the system during this cycle. We assume 
the following form of the variation: $f(t)=(1-\sqrt{\sin(A t/\tau)})$, where $\tau\equiv \pi/J_1[0]$ defines the natural time scale in the box 
in the absence of the barrier. Note that one cycle happens within the period $\pi/(A \tau)$.  To supress the dynamics between different wells, we put 
$\alpha=50$ (see Fig.~\ref{fig:ratio}). We present our findings in Fig.~\ref{fig:ferr}, showing the probabilities of different configurations, 
i.e. $|a_i|^2$ for different $A$. Note that these probabilities after one cycle depend strongly
on $A$ which provides a way for state preparation. For example let us 
take a look at the case with $A=0.8$ (panel c)). We see that for such a driving mode one ends up in $0.996\%$ of all cases in 
$\downarrow\downarrow\uparrow\uparrow$ which can be seen as the exchange of the pairs. 
 Note, that if we had plotted the total density before and after the cycle for every configuration from Fig. \ref{fig:ferr}, then we would have obtained the same result due to the adiabaticity of particle motion. Nevertheless, the spin configurations are profoundly different which highlights the separation of the spin and particle dynamics that we have derived.  It is worthwhile noting that for even slower change of the potential with the time scale much larger than $g_f$ the dynamics in the both spin and charge sectors will be adiabatic.

{\it Conclusions}. -- In this paper, we discuss a time dependent spin chain which is realized with strongly interacting atoms in a time-dependent confinement in 1D. First, we outline a mapping onto a spin chain for one impurity in a Fermi sea of majority 
particles. This mapping can be trivially extended to more impurity particles or other multicomponent strongly coupled systems. 
Next, we use a time-dependent harmonic oscillator potential with a weak stationary magnetic field and an infinite square well potential with a 
time-dependent barrier to illustrate some basic properties of the spin dynamics in such systems. In particular, we show that in the former 
case by changing the trapping potential one can drive a system to a spin segregated state. For the latter case, 
we demonstrate the possibility of a state preparation and manipulation by proper changing the shape of the trapping potential.

A major goal of cold atomic gas research is to reach the regime where 
quantum magnetism can be studied and a number of pioneering experiments
have already been reported \cite{anderlini2007,folling2007,trotzky2008}. 
In particular, the superexchange of two spins has been observed 
in Ref.~\cite{trotzky2008} and it was shown that the lattice spin model 
limit of the Bose-Hubbard model \cite{kuklov2003,duan2003} could accurately describe the data.
While limited to strong cooupling and 1D,
the approach  described here goes beyond those models as it can fully incorporate
the shape of any (time-dependent) potential, circumventing any need for making a 
lattice approximation. 
Our approach therefore ideally complements the lattice approximation as a tool to simulate and study spin dynamics.
This allows us to address the 
dynamical evolution of general $N$-body exchanges in arbitrary potentials in the
strongly interacting limit for both fermionic and bosonic atoms. 
Our theory may therefore be relevant for using
exchange interactions to generate multiparticle entanglement \cite{briegel2001,mandel2003}
and building robust quantum gates \cite{burkard1999,petta2005} for
use in quantum communication \cite{korzh2015}, computation, and information \cite{nielsen2010}.

\begin{acknowledgments}

A. G. V. and N. T. Z. would like to thank A. S. Dehkharghani, A. S. Jensen, D.
Fedorov, C. Forss{\'e}n, E. J. Lindgren, O. V. Marchukov,  D. Petrosyan, J. Rotureau and M. Valiente for
collaboration on strongly interacting 1D systems.  
We acknowledge discussions with the participants of the 595th WE-Heraeus Seminar "Cold Atoms meet QFT".
This work was supported in part by Helmholtz Association 
under contract HA216/EMMI, by the BMBF (grant 06BN9006), and by the Danish Council for Independent Research DFF Natural Sciences

\end{acknowledgments}

\appendix

\widetext

\section{Supplemental Material for "Simulation of Time-dependent Heisenberg Models in 1D"}

Here we outline the derivation of Eq. (5) of the main text. For this we first insert the wave function $\Psi=\phi+\frac{1}{g_f}f$ in the Schr{\"o}dinger equation. 
Next we make a projection onto a specific ordering by integrating the equation with $\chi_j\Phi^*$. This procedure yields
\begin{equation}
i \langle \chi_j  \Phi|\Phi \rangle \frac{\mathrm{d}a_j(t)}{\mathrm{d}t}+i a_j \langle \chi_j \Phi|\frac{\partial \Phi}{\partial t} \rangle + \frac{i}{g_f}\langle \chi_j \Phi|\frac{\partial f}{\partial t} \rangle = \langle \chi_j \Phi|H|\phi \rangle +\frac{1}{g_f}\langle \chi_j \Phi|H|f \rangle \; .
\label{eq:suppl}
\end{equation}
To proceed, we notice that $H\chi_j\Phi=i\frac{\partial \chi_j \Phi}{\partial t}$ 
everywhere except at the points where the particles meet. We also notice that due to a non-smooth behaviour close to these points $H\chi_j \Phi$ yields a Dirac delta function. These observations allow us to conclude that   
\begin{equation}
\langle \chi_j \Phi|H|\phi \rangle = i a_j(t) \langle \chi_j \Phi |\frac{\partial \Phi}{\partial t} \rangle.
\end{equation}
Next we turn our attention to the $\langle \chi_j \Phi|H|f \rangle$ term in Eq. (\ref{eq:suppl}) which, for convenience, we rewrite as an integral over the 
configuration $y_1<y_2<...<y_{j-1}<x<y_j<...<y_N$,
\begin{equation}
\langle \chi_j \Phi|H|f \rangle = N! \int_{-\infty}^\infty \mathrm{d}y_1 \int_{y_1}^\infty \mathrm{d} y_2 ... \int_{y_{i-1}}^\infty \mathrm{d}x \int_{x}^\infty\mathrm{d}y_j...\int_{y_{N-1}}^\infty \mathrm{d}y_N  \mathrm{d}x ...\mathrm{d}y_N
\Phi^* H f.
\end{equation}
Our next steps are two integrations by parts. This will yield some boundary terms and the integral with $f$ and $\Phi$ exchanged. Notice that there will be two types of boundary terms: $i)$ with $y_l=y_k$, and $ii)$ with $x=y_{j-1}$ or $x=y_{j}$. The former terms vanish due to the fermionic nature of the majority particles. The latter, however, should be properly taken into account,
\begin{align}
\frac{\langle \chi_j \Phi|H|f \rangle}{N!} = -\frac{1}{2}\lim_{\epsilon\to0}\int_{y_1<y_2<...<y_{j-1}<x<y_j<...<y_N}\mathrm{d}x \mathrm{d}y_1...\mathrm{d}y_N \bigg( [\delta(x-y_j+\epsilon) - \delta(x-y_{j-1}-\epsilon)]\left(\Phi^* \frac{\partial f}{\partial x}-f\frac{\partial\Phi^*}{\partial x}\right)  \nonumber \\ + \delta(x-y_{j-1}-\epsilon) \left(\Phi^* \frac{\partial f}{\partial y_{j-1}}-f \frac{\partial \Phi^*}{\partial y_{j-1}}\right)  - \delta(x-y_j+\epsilon) \left( \Phi^*\frac{\partial f}{\partial y_j} - f\frac{\partial \Phi^*}{\partial y_j} \right) -2 f H \Phi^* \bigg),
\end{align}
where all but the last terms under the integral sign are the boundary terms. Thus, we have
\begin{equation}
\langle \chi_j \Phi|H|f \rangle =-i\langle f\chi_j| \frac{\partial \Phi}{\partial t} \rangle ^* + \mathrm{boundary\;terms}\;. 
\end{equation}
Next, we collect the expressions just derived and write the equation for $a_j(t)$
\begin{equation}
i\frac{\mathrm{d}a_j(t)}{\mathrm{d}t}=\frac{1}{g_f}\frac{\mathrm{boundary\;terms}}{ \langle \chi_j  \Phi|\Phi \rangle },
\label{eq:ajbound}
\end{equation}
where we used that $\frac{\mathrm{d} { \langle \chi_j  \Phi|f \rangle }}{\mathrm{d}t}=0$ by construction. It is important to notice that Eq. (\ref{eq:ajbound}) is general and does not rely on the assumption that $1/g_f$ is small. However, as we show below this assumption makes the derived expression very useful. Let us now focus on the boundary terms, 
\begin{align}
\mathrm{boundary\;terms} = &-\frac{N!}{2}\lim_{\epsilon\to0}\int_{y_1<y_2<...<y_{j-1}<x<y_j<...<y_N}\mathrm{d}x \mathrm{d}y_1...\mathrm{d}y_N f \times \nonumber \\ 
&\bigg( \delta(x-y_j+\epsilon)\left(\frac{\partial \Phi^*}{\partial y_j}-\frac{\partial\Phi^*}{\partial x}\right) + \delta(x-y_{j-1}-\epsilon) \left( \frac{\partial \Phi^*}{\partial x}- \frac{\partial \Phi^*}{\partial y_{j-1}}\right) \bigg) \nonumber =\\
 &N!\int_{y_1<y_2<...<y_{j-1}<x<y_j<...<y_N}\mathrm{d}x \mathrm{d}y_1...\mathrm{d}y_N f 
\bigg( \delta(x-y_j) - \delta(x-y_{j-1}) \bigg)\frac{\partial \Phi^*}{\partial x}.
\label{eq:bound_terms}
\end{align}
To obtain $f$ at the points where the particles meet, we use the boundary conditions from the main text. This yields
\begin{equation}
f(x=y_i)=(a_{i+1}-a_i)\frac{\partial \Phi}{\partial x}\bigg|_{x=y_i}+\frac{1}{2g_f}\left[\left(\frac{\partial f}{\partial x}-\frac{\partial f}{\partial y_i}\right)_{x=y_i^+}-
\left(\frac{\partial f}{\partial x}-\frac{\partial f}{\partial y_i}\right)_{x=y_i^-}\right].
\label{eq:f_part_meet}
\end{equation}
Now we make the assumption that $1/g_f\to0$, which implies that $f(x=y_i)\simeq(a_{i+1}-a_i)\frac{\partial \Phi}{\partial x}\big|_{x=y_i}$. Inserting this result in Eqs. (\ref{eq:bound_terms}) and (\ref{eq:ajbound}), we arrive at the desired expression.  It should be noted that in our derivations we assume that $g_f$ sets the largest energy scale of the problem. Therefore, if the change of the trap is such that $g_f$ becomes of the order of the other energy scales then the treatment above is not valid, namely we cannot neglect the second term on the right-hand-side of Eq. (\ref{eq:f_part_meet}). For example this can happen if we increase the density of the system, by squeezing the trap, which necessarily increases the kinetic energy. Another instance is a periodic driving in the parametric resonance region, which pumps in energy in the system.

\bibliographystyle{apsrev4-1}
\bibliography{bib}

\begin{thebibliography}{42}%
\makeatletter
\providecommand \@ifxundefined [1]{%
 \@ifx{#1\undefined}
}%
\providecommand \@ifnum [1]{%
 \ifnum #1\expandafter \@firstoftwo
 \else \expandafter \@secondoftwo
 \fi
}%
\providecommand \@ifx [1]{%
 \ifx #1\expandafter \@firstoftwo
 \else \expandafter \@secondoftwo
 \fi
}%
\providecommand \natexlab [1]{#1}%
\providecommand \enquote  [1]{``#1''}%
\providecommand \bibnamefont  [1]{#1}%
\providecommand \bibfnamefont [1]{#1}%
\providecommand \citenamefont [1]{#1}%
\providecommand \href@noop [0]{\@secondoftwo}%
\providecommand \href [0]{\begingroup \@sanitize@url \@href}%
\providecommand \@href[1]{\@@startlink{#1}\@@href}%
\providecommand \@@href[1]{\endgroup#1\@@endlink}%
\providecommand \@sanitize@url [0]{\catcode `\\12\catcode `\$12\catcode
  `\&12\catcode `\#12\catcode `\^12\catcode `\_12\catcode `\%12\relax}%
\providecommand \@@startlink[1]{}%
\providecommand \@@endlink[0]{}%
\providecommand \url  [0]{\begingroup\@sanitize@url \@url }%
\providecommand \@url [1]{\endgroup\@href {#1}{\urlprefix }}%
\providecommand \urlprefix  [0]{URL }%
\providecommand \Eprint [0]{\href }%
\providecommand \doibase [0]{http://dx.doi.org/}%
\providecommand \selectlanguage [0]{\@gobble}%
\providecommand \bibinfo  [0]{\@secondoftwo}%
\providecommand \bibfield  [0]{\@secondoftwo}%
\providecommand \translation [1]{[#1]}%
\providecommand \BibitemOpen [0]{}%
\providecommand \bibitemStop [0]{}%
\providecommand \bibitemNoStop [0]{.\EOS\space}%
\providecommand \EOS [0]{\spacefactor3000\relax}%
\providecommand \BibitemShut  [1]{\csname bibitem#1\endcsname}%
\let\auto@bib@innerbib\@empty
\bibitem [{\citenamefont {Cazalilla}\ \emph {et~al.}(2011)\citenamefont
  {Cazalilla}, \citenamefont {Citro}, \citenamefont {Giamarchi}, \citenamefont
  {Orignac},\ and\ \citenamefont {Rigol}}]{rigol2011}%
  \BibitemOpen
  \bibfield  {author} {\bibinfo {author} {\bibfnamefont {M.~A.}\ \bibnamefont
  {Cazalilla}}, \bibinfo {author} {\bibfnamefont {R.}~\bibnamefont {Citro}},
  \bibinfo {author} {\bibfnamefont {T.}~\bibnamefont {Giamarchi}}, \bibinfo
  {author} {\bibfnamefont {E.}~\bibnamefont {Orignac}}, \ and\ \bibinfo
  {author} {\bibfnamefont {M.}~\bibnamefont {Rigol}},\ }\href {\doibase
  10.1103/RevModPhys.83.1405} {\bibfield  {journal} {\bibinfo  {journal} {Rev.
  Mod. Phys.}\ }\textbf {\bibinfo {volume} {83}},\ \bibinfo {pages} {1405}
  (\bibinfo {year} {2011})}\BibitemShut {NoStop}%
\bibitem [{\citenamefont {Girardeau}(1960)}]{girardeau1960}%
  \BibitemOpen
  \bibfield  {author} {\bibinfo {author} {\bibfnamefont {M.}~\bibnamefont
  {Girardeau}},\ }\href {\doibase http://dx.doi.org/10.1063/1.1703687}
  {\bibfield  {journal} {\bibinfo  {journal} {Journal of Mathematical Physics}\
  }\textbf {\bibinfo {volume} {1}},\ \bibinfo {pages} {516} (\bibinfo {year}
  {1960})}\BibitemShut {NoStop}%
\bibitem [{\citenamefont {Paredes}\ \emph {et~al.}(2004)\citenamefont
  {Paredes}, \citenamefont {Widera}, \citenamefont {Murg}, \citenamefont
  {Mandel}, \citenamefont {Folling}, \citenamefont {Cirac}, \citenamefont
  {Shlyapnikov}, \citenamefont {Hansch},\ and\ \citenamefont {Bloch}}]{tg2004}%
  \BibitemOpen
  \bibfield  {author} {\bibinfo {author} {\bibfnamefont {B.}~\bibnamefont
  {Paredes}}, \bibinfo {author} {\bibfnamefont {A.}~\bibnamefont {Widera}},
  \bibinfo {author} {\bibfnamefont {V.}~\bibnamefont {Murg}}, \bibinfo {author}
  {\bibfnamefont {O.}~\bibnamefont {Mandel}}, \bibinfo {author} {\bibfnamefont
  {S.}~\bibnamefont {Folling}}, \bibinfo {author} {\bibfnamefont
  {I.}~\bibnamefont {Cirac}}, \bibinfo {author} {\bibfnamefont {G.~V.}\
  \bibnamefont {Shlyapnikov}}, \bibinfo {author} {\bibfnamefont {T.~W.}\
  \bibnamefont {Hansch}}, \ and\ \bibinfo {author} {\bibfnamefont
  {I.}~\bibnamefont {Bloch}},\ }\href {http://dx.doi.org/10.1038/nature02530}
  {\bibfield  {journal} {\bibinfo  {journal} {Nature}\ }\textbf {\bibinfo
  {volume} {429}},\ \bibinfo {pages} {277} (\bibinfo {year}
  {2004})}\BibitemShut {NoStop}%
\bibitem [{\citenamefont {Kinoshita}\ \emph {et~al.}(2004)\citenamefont
  {Kinoshita}, \citenamefont {Wenger},\ and\ \citenamefont
  {Weiss}}]{Kinoshita20082004}%
  \BibitemOpen
  \bibfield  {author} {\bibinfo {author} {\bibfnamefont {T.}~\bibnamefont
  {Kinoshita}}, \bibinfo {author} {\bibfnamefont {T.}~\bibnamefont {Wenger}}, \
  and\ \bibinfo {author} {\bibfnamefont {D.~S.}\ \bibnamefont {Weiss}},\ }\href
  {\doibase 10.1126/science.1100700} {\bibfield  {journal} {\bibinfo  {journal}
  {Science}\ }\textbf {\bibinfo {volume} {305}},\ \bibinfo {pages} {1125}
  (\bibinfo {year} {2004})}\BibitemShut {NoStop}%
\bibitem [{\citenamefont {Z\"urn}\ \emph {et~al.}(2012)\citenamefont {Z\"urn},
  \citenamefont {Serwane}, \citenamefont {Lompe}, \citenamefont {Wenz},
  \citenamefont {Ries}, \citenamefont {Bohn},\ and\ \citenamefont
  {Jochim}}]{jochim2012}%
  \BibitemOpen
  \bibfield  {author} {\bibinfo {author} {\bibfnamefont {G.}~\bibnamefont
  {Z\"urn}}, \bibinfo {author} {\bibfnamefont {F.}~\bibnamefont {Serwane}},
  \bibinfo {author} {\bibfnamefont {T.}~\bibnamefont {Lompe}}, \bibinfo
  {author} {\bibfnamefont {A.~N.}\ \bibnamefont {Wenz}}, \bibinfo {author}
  {\bibfnamefont {M.~G.}\ \bibnamefont {Ries}}, \bibinfo {author}
  {\bibfnamefont {J.~E.}\ \bibnamefont {Bohn}}, \ and\ \bibinfo {author}
  {\bibfnamefont {S.}~\bibnamefont {Jochim}},\ }\href {\doibase
  10.1103/PhysRevLett.108.075303} {\bibfield  {journal} {\bibinfo  {journal}
  {Phys. Rev. Lett.}\ }\textbf {\bibinfo {volume} {108}},\ \bibinfo {pages}
  {075303} (\bibinfo {year} {2012})}\BibitemShut {NoStop}%
\bibitem [{\citenamefont {Volosniev}\ \emph {et~al.}(2014)\citenamefont
  {Volosniev}, \citenamefont {Fedorov}, \citenamefont {Jensen},\ and\
  \citenamefont {Zinner}}]{volosniev2014}%
  \BibitemOpen
  \bibfield  {author} {\bibinfo {author} {\bibfnamefont {A.~G.}\ \bibnamefont
  {Volosniev}}, \bibinfo {author} {\bibfnamefont {D.~V.}\ \bibnamefont
  {Fedorov}}, \bibinfo {author} {\bibfnamefont {M.}~\bibnamefont {Jensen},
  \bibfnamefont {A.~S.~Valiente}}, \ and\ \bibinfo {author} {\bibfnamefont
  {N.~T.}\ \bibnamefont {Zinner}},\ }\href {\doibase
  http://dx.doi.org/10.1038/ncomms6300} {\bibfield  {journal} {\bibinfo
  {journal} {Nat. Commun.}\ }\textbf {\bibinfo {volume} {5}},\ \bibinfo {pages}
  {5300} (\bibinfo {year} {2014})}\BibitemShut {NoStop}%
\bibitem [{\citenamefont {Deuretzbacher}\ \emph {et~al.}(2014)\citenamefont
  {Deuretzbacher}, \citenamefont {Becker}, \citenamefont {Bjerlin},
  \citenamefont {Reimann},\ and\ \citenamefont {Santos}}]{deuret2014}%
  \BibitemOpen
  \bibfield  {author} {\bibinfo {author} {\bibfnamefont {F.}~\bibnamefont
  {Deuretzbacher}}, \bibinfo {author} {\bibfnamefont {D.}~\bibnamefont
  {Becker}}, \bibinfo {author} {\bibfnamefont {J.}~\bibnamefont {Bjerlin}},
  \bibinfo {author} {\bibfnamefont {S.~M.}\ \bibnamefont {Reimann}}, \ and\
  \bibinfo {author} {\bibfnamefont {L.}~\bibnamefont {Santos}},\ }\href
  {\doibase 10.1103/PhysRevA.90.013611} {\bibfield  {journal} {\bibinfo
  {journal} {Phys. Rev. A}\ }\textbf {\bibinfo {volume} {90}},\ \bibinfo
  {pages} {013611} (\bibinfo {year} {2014})}\BibitemShut {NoStop}%
\bibitem [{\citenamefont {Volosniev}\ \emph
  {et~al.}(2015{\natexlab{a}})\citenamefont {Volosniev}, \citenamefont
  {Petrosyan}, \citenamefont {Valiente}, \citenamefont {Fedorov}, \citenamefont
  {Jensen},\ and\ \citenamefont {Zinner}}]{volosniev2015}%
  \BibitemOpen
  \bibfield  {author} {\bibinfo {author} {\bibfnamefont {A.~G.}\ \bibnamefont
  {Volosniev}}, \bibinfo {author} {\bibfnamefont {D.}~\bibnamefont
  {Petrosyan}}, \bibinfo {author} {\bibfnamefont {M.}~\bibnamefont {Valiente}},
  \bibinfo {author} {\bibfnamefont {D.~V.}\ \bibnamefont {Fedorov}}, \bibinfo
  {author} {\bibfnamefont {A.~S.}\ \bibnamefont {Jensen}}, \ and\ \bibinfo
  {author} {\bibfnamefont {N.~T.}\ \bibnamefont {Zinner}},\ }\href {\doibase
  10.1103/PhysRevA.91.023620} {\bibfield  {journal} {\bibinfo  {journal} {Phys.
  Rev. A}\ }\textbf {\bibinfo {volume} {91}},\ \bibinfo {pages} {023620}
  (\bibinfo {year} {2015}{\natexlab{a}})}\BibitemShut {NoStop}%
\bibitem [{\citenamefont {Levinsen}\ \emph {et~al.}(2015)\citenamefont
  {Levinsen}, \citenamefont {Massignan}, \citenamefont {Bruun},\ and\
  \citenamefont {Parish}}]{levinsen2014}%
  \BibitemOpen
  \bibfield  {author} {\bibinfo {author} {\bibfnamefont {J.}~\bibnamefont
  {Levinsen}}, \bibinfo {author} {\bibfnamefont {P.}~\bibnamefont {Massignan}},
  \bibinfo {author} {\bibfnamefont {G.~M.}\ \bibnamefont {Bruun}}, \ and\
  \bibinfo {author} {\bibfnamefont {M.~M.}\ \bibnamefont {Parish}},\
  }\href@noop {} {\bibfield  {journal} {\bibinfo  {journal} {Science Advances}\
  }\textbf {\bibinfo {volume} {1}},\ \bibinfo {pages} {e1500197} (\bibinfo
  {year} {2015})}\BibitemShut {NoStop}%
\bibitem [{\citenamefont {Serwane}\ \emph {et~al.}(2011)\citenamefont
  {Serwane}, \citenamefont {Z{\"u}rn}, \citenamefont {Lompe}, \citenamefont
  {Ottenstein}, \citenamefont {Wenz},\ and\ \citenamefont
  {Jochim}}]{serwane2011}%
  \BibitemOpen
  \bibfield  {author} {\bibinfo {author} {\bibfnamefont {F.}~\bibnamefont
  {Serwane}}, \bibinfo {author} {\bibfnamefont {G.}~\bibnamefont {Z{\"u}rn}},
  \bibinfo {author} {\bibfnamefont {T.}~\bibnamefont {Lompe}}, \bibinfo
  {author} {\bibfnamefont {T.~B.}\ \bibnamefont {Ottenstein}}, \bibinfo
  {author} {\bibfnamefont {A.~N.}\ \bibnamefont {Wenz}}, \ and\ \bibinfo
  {author} {\bibfnamefont {S.}~\bibnamefont {Jochim}},\ }\href {\doibase
  10.1126/science.1201351} {\bibfield  {journal} {\bibinfo  {journal}
  {Science}\ }\textbf {\bibinfo {volume} {332}},\ \bibinfo {pages} {336}
  (\bibinfo {year} {2011})}\BibitemShut {NoStop}%
\bibitem [{\citenamefont {Pagano}\ \emph {et~al.}(2014)\citenamefont {Pagano},
  \citenamefont {Mancini}, \citenamefont {Cappellini}, \citenamefont
  {Lombardi}, \citenamefont {Schafer}, \citenamefont {Hu}, \citenamefont {Liu},
  \citenamefont {Catani}, \citenamefont {Sias}, \citenamefont {Inguscio},\ and\
  \citenamefont {Fallani}}]{pagano2014}%
  \BibitemOpen
  \bibfield  {author} {\bibinfo {author} {\bibfnamefont {G.}~\bibnamefont
  {Pagano}}, \bibinfo {author} {\bibfnamefont {M.}~\bibnamefont {Mancini}},
  \bibinfo {author} {\bibfnamefont {G.}~\bibnamefont {Cappellini}}, \bibinfo
  {author} {\bibfnamefont {P.}~\bibnamefont {Lombardi}}, \bibinfo {author}
  {\bibfnamefont {F.}~\bibnamefont {Schafer}}, \bibinfo {author} {\bibfnamefont
  {H.}~\bibnamefont {Hu}}, \bibinfo {author} {\bibfnamefont {X.-J.}\
  \bibnamefont {Liu}}, \bibinfo {author} {\bibfnamefont {J.}~\bibnamefont
  {Catani}}, \bibinfo {author} {\bibfnamefont {C.}~\bibnamefont {Sias}},
  \bibinfo {author} {\bibfnamefont {M.}~\bibnamefont {Inguscio}}, \ and\
  \bibinfo {author} {\bibfnamefont {L.}~\bibnamefont {Fallani}},\ }\href
  {http://dx.doi.org/10.1038/nphys2878} {\bibfield  {journal} {\bibinfo
  {journal} {Nat. Phys.}\ }\textbf {\bibinfo {volume} {10}},\ \bibinfo {pages}
  {198} (\bibinfo {year} {2014})}\BibitemShut {NoStop}%
\bibitem [{\citenamefont {Murmann}\ \emph {et~al.}(2015)\citenamefont
  {Murmann}, \citenamefont {Bergschneider}, \citenamefont {Klinkhamer},
  \citenamefont {Z\"urn}, \citenamefont {Lompe},\ and\ \citenamefont
  {Jochim}}]{jochim2014}%
  \BibitemOpen
  \bibfield  {author} {\bibinfo {author} {\bibfnamefont {S.}~\bibnamefont
  {Murmann}}, \bibinfo {author} {\bibfnamefont {A.}~\bibnamefont
  {Bergschneider}}, \bibinfo {author} {\bibfnamefont {V.~M.}\ \bibnamefont
  {Klinkhamer}}, \bibinfo {author} {\bibfnamefont {G.}~\bibnamefont {Z\"urn}},
  \bibinfo {author} {\bibfnamefont {T.}~\bibnamefont {Lompe}}, \ and\ \bibinfo
  {author} {\bibfnamefont {S.}~\bibnamefont {Jochim}},\ }\href {\doibase
  10.1103/PhysRevLett.114.080402} {\bibfield  {journal} {\bibinfo  {journal}
  {Phys. Rev. Lett.}\ }\textbf {\bibinfo {volume} {114}},\ \bibinfo {pages}
  {080402} (\bibinfo {year} {2015})}\BibitemShut {NoStop}%
\bibitem [{\citenamefont {Bose}(2003)}]{bose2003}%
  \BibitemOpen
  \bibfield  {author} {\bibinfo {author} {\bibfnamefont {S.}~\bibnamefont
  {Bose}},\ }\href {\doibase 10.1103/PhysRevLett.91.207901} {\bibfield
  {journal} {\bibinfo  {journal} {Phys. Rev. Lett.}\ }\textbf {\bibinfo
  {volume} {91}},\ \bibinfo {pages} {207901} (\bibinfo {year}
  {2003})}\BibitemShut {NoStop}%
\bibitem [{\citenamefont {Girardeau}\ and\ \citenamefont
  {Wright}(2000{\natexlab{a}})}]{wright2000}%
  \BibitemOpen
  \bibfield  {author} {\bibinfo {author} {\bibfnamefont {M.~D.}\ \bibnamefont
  {Girardeau}}\ and\ \bibinfo {author} {\bibfnamefont {E.~M.}\ \bibnamefont
  {Wright}},\ }\href {\doibase 10.1103/PhysRevLett.84.5691} {\bibfield
  {journal} {\bibinfo  {journal} {Phys. Rev. Lett.}\ }\textbf {\bibinfo
  {volume} {84}},\ \bibinfo {pages} {5691} (\bibinfo {year}
  {2000}{\natexlab{a}})}\BibitemShut {NoStop}%
\bibitem [{\citenamefont {Girardeau}\ and\ \citenamefont
  {Wright}(2000{\natexlab{b}})}]{girardeau2000}%
  \BibitemOpen
  \bibfield  {author} {\bibinfo {author} {\bibfnamefont {M.~D.}\ \bibnamefont
  {Girardeau}}\ and\ \bibinfo {author} {\bibfnamefont {E.~M.}\ \bibnamefont
  {Wright}},\ }\href {\doibase 10.1103/PhysRevLett.84.5239} {\bibfield
  {journal} {\bibinfo  {journal} {Phys. Rev. Lett.}\ }\textbf {\bibinfo
  {volume} {84}},\ \bibinfo {pages} {5239} (\bibinfo {year}
  {2000}{\natexlab{b}})}\BibitemShut {NoStop}%
\bibitem [{Note1()}]{Note1}%
  \BibitemOpen
  \bibinfo {note} {Recent studies have considered time-dependent driving
  starting from the lattice approximation \cite {itin2014}. However we note
  that our formalism does not require one to make a lattice model
  approximation.}\BibitemShut {Stop}%
\bibitem [{\citenamefont {Lyakhov}\ and\ \citenamefont
  {Bruder}(2006)}]{Bruder2006}%
  \BibitemOpen
  \bibfield  {author} {\bibinfo {author} {\bibfnamefont {A.~O.}\ \bibnamefont
  {Lyakhov}}\ and\ \bibinfo {author} {\bibfnamefont {C.}~\bibnamefont
  {Bruder}},\ }\href {\doibase 10.1103/PhysRevB.74.235303} {\bibfield
  {journal} {\bibinfo  {journal} {Phys. Rev. B}\ }\textbf {\bibinfo {volume}
  {74}},\ \bibinfo {pages} {235303} (\bibinfo {year} {2006})}\BibitemShut
  {NoStop}%
\bibitem [{\citenamefont {Galve}\ \emph {et~al.}(2009)\citenamefont {Galve},
  \citenamefont {Zueco}, \citenamefont {Kohler}, \citenamefont {Lutz},\ and\
  \citenamefont {H\"anggi}}]{Galve2009}%
  \BibitemOpen
  \bibfield  {author} {\bibinfo {author} {\bibfnamefont {F.}~\bibnamefont
  {Galve}}, \bibinfo {author} {\bibfnamefont {D.}~\bibnamefont {Zueco}},
  \bibinfo {author} {\bibfnamefont {S.}~\bibnamefont {Kohler}}, \bibinfo
  {author} {\bibfnamefont {E.}~\bibnamefont {Lutz}}, \ and\ \bibinfo {author}
  {\bibfnamefont {P.}~\bibnamefont {H\"anggi}},\ }\href {\doibase
  10.1103/PhysRevA.79.032332} {\bibfield  {journal} {\bibinfo  {journal} {Phys.
  Rev. A}\ }\textbf {\bibinfo {volume} {79}},\ \bibinfo {pages} {032332}
  (\bibinfo {year} {2009})}\BibitemShut {NoStop}%
\bibitem [{\citenamefont {McGuire}(1965)}]{McGuire1965}%
  \BibitemOpen
  \bibfield  {author} {\bibinfo {author} {\bibfnamefont {J.~B.}\ \bibnamefont
  {McGuire}},\ }\href {\doibase http://dx.doi.org/10.1063/1.1704291} {\bibfield
   {journal} {\bibinfo  {journal} {Journal of Mathematical Physics}\ }\textbf
  {\bibinfo {volume} {6}},\ \bibinfo {pages} {432} (\bibinfo {year}
  {1965})}\BibitemShut {NoStop}%
\bibitem [{\citenamefont {Volosniev}\ \emph
  {et~al.}(2015{\natexlab{b}})\citenamefont {Volosniev}, \citenamefont
  {Fedorov}, \citenamefont {Jensen},\ and\ \citenamefont
  {Zinner}}]{volosniev2014a}%
  \BibitemOpen
  \bibfield  {author} {\bibinfo {author} {\bibfnamefont {A.~G.}\ \bibnamefont
  {Volosniev}}, \bibinfo {author} {\bibfnamefont {D.~V.}\ \bibnamefont
  {Fedorov}}, \bibinfo {author} {\bibfnamefont {A.~S.}\ \bibnamefont {Jensen}},
  \ and\ \bibinfo {author} {\bibfnamefont {N.~T.}\ \bibnamefont {Zinner}},\
  }\href@noop {} {\bibfield  {journal} {\bibinfo  {journal} {Eur. Phys. J.
  Special Topics}\ }\textbf {\bibinfo {volume} {224}},\ \bibinfo {pages} {585}
  (\bibinfo {year} {2015}{\natexlab{b}})}\BibitemShut {NoStop}%
\bibitem [{\citenamefont {Gharashi}\ \emph {et~al.}(2015)\citenamefont
  {Gharashi}, \citenamefont {Yin}, \citenamefont {Yan},\ and\ \citenamefont
  {Blume}}]{blume2015}%
  \BibitemOpen
  \bibfield  {author} {\bibinfo {author} {\bibfnamefont {S.~E.}\ \bibnamefont
  {Gharashi}}, \bibinfo {author} {\bibfnamefont {X.~Y.}\ \bibnamefont {Yin}},
  \bibinfo {author} {\bibfnamefont {Y.}~\bibnamefont {Yan}}, \ and\ \bibinfo
  {author} {\bibfnamefont {D.}~\bibnamefont {Blume}},\ }\href {\doibase
  10.1103/PhysRevA.91.013620} {\bibfield  {journal} {\bibinfo  {journal} {Phys.
  Rev. A}\ }\textbf {\bibinfo {volume} {91}},\ \bibinfo {pages} {013620}
  (\bibinfo {year} {2015})}\BibitemShut {NoStop}%
\bibitem [{sup()}]{supp}%
  \BibitemOpen
  \href@noop {} {\bibinfo  {journal} {See Supplemental Material}\ }\BibitemShut
  {NoStop}%
\bibitem [{\citenamefont {Giamarchi}(2004)}]{giamarchi-book}%
  \BibitemOpen
\bibfield  {journal} {  }\bibfield  {author} {\bibinfo {author} {\bibfnamefont
  {T.}~\bibnamefont {Giamarchi}},\ }\href@noop {} {\emph {\bibinfo {title}
  {Quantum Physics in One Dimension}}}\ (\bibinfo  {publisher} {Clarendon
  Press, Oxford},\ \bibinfo {year} {2004})\BibitemShut {NoStop}%
\bibitem [{\citenamefont {Husimi}(1953)}]{Husimi1953}%
  \BibitemOpen
  \bibfield  {author} {\bibinfo {author} {\bibfnamefont {K.}~\bibnamefont
  {Husimi}},\ }\href {\doibase 10.1143/ptp/9.4.381} {\bibfield  {journal}
  {\bibinfo  {journal} {Progress of Theoretical Physics}\ }\textbf {\bibinfo
  {volume} {9}},\ \bibinfo {pages} {381} (\bibinfo {year} {1953})}\BibitemShut
  {NoStop}%
\bibitem [{\citenamefont {Popov}\ and\ \citenamefont
  {Perelomov}(1969)}]{perelomov1969}%
  \BibitemOpen
  \bibfield  {author} {\bibinfo {author} {\bibfnamefont {V.~S.}\ \bibnamefont
  {Popov}}\ and\ \bibinfo {author} {\bibfnamefont {A.~M.}\ \bibnamefont
  {Perelomov}},\ }\href@noop {} {\bibfield  {journal} {\bibinfo  {journal}
  {JETP}\ }\textbf {\bibinfo {volume} {29}},\ \bibinfo {pages} {738} (\bibinfo
  {year} {1969})}\BibitemShut {NoStop}%
\bibitem [{\citenamefont {Castin}(2004)}]{castin2004}%
  \BibitemOpen
  \bibfield  {author} {\bibinfo {author} {\bibfnamefont {Y.}~\bibnamefont
  {Castin}},\ }\href {\doibase http://dx.doi.org/10.1016/j.crhy.2004.03.017}
  {\bibfield  {journal} {\bibinfo  {journal} {Comptes Rendus Physique}\
  }\textbf {\bibinfo {volume} {5}},\ \bibinfo {pages} {407 } (\bibinfo {year}
  {2004})}\BibitemShut {NoStop}%
\bibitem [{\citenamefont {Moroz}(2012)}]{moroz2012}%
  \BibitemOpen
  \bibfield  {author} {\bibinfo {author} {\bibfnamefont {S.}~\bibnamefont
  {Moroz}},\ }\href {\doibase 10.1103/PhysRevA.86.011601} {\bibfield  {journal}
  {\bibinfo  {journal} {Phys. Rev. A}\ }\textbf {\bibinfo {volume} {86}},\
  \bibinfo {pages} {011601} (\bibinfo {year} {2012})}\BibitemShut {NoStop}%
\bibitem [{\citenamefont {Ebert}\ \emph {et~al.}(2015)\citenamefont {Ebert},
  \citenamefont {Volosniev},\ and\ \citenamefont {Hammer}}]{Ebert:2015gko}%
  \BibitemOpen
  \bibfield  {author} {\bibinfo {author} {\bibfnamefont {M.}~\bibnamefont
  {Ebert}}, \bibinfo {author} {\bibfnamefont {A.}~\bibnamefont {Volosniev}}, \
  and\ \bibinfo {author} {\bibfnamefont {H.~W.}\ \bibnamefont {Hammer}},\
  }\href@noop {} {\  (\bibinfo {year} {2015})},\ \Eprint
  {http://arxiv.org/abs/1512.06628} {arXiv:1512.06628 [cond-mat.quant-gas]}
  \BibitemShut {NoStop}%
\bibitem [{Note2()}]{Note2}%
  \BibitemOpen
  \bibinfo {note} {Notice, that the resulting Hamiltonian can be
  straighforwardly generalized to more particles}\BibitemShut {NoStop}%
\bibitem [{\citenamefont {Cui}\ and\ \citenamefont {Ho}(2014)}]{cui2015}%
  \BibitemOpen
  \bibfield  {author} {\bibinfo {author} {\bibfnamefont {X.}~\bibnamefont
  {Cui}}\ and\ \bibinfo {author} {\bibfnamefont {T.-L.}\ \bibnamefont {Ho}},\
  }\href {\doibase 10.1103/PhysRevA.89.023611} {\bibfield  {journal} {\bibinfo
  {journal} {Phys. Rev. A}\ }\textbf {\bibinfo {volume} {89}},\ \bibinfo
  {pages} {023611} (\bibinfo {year} {2014})}\BibitemShut {NoStop}%
\bibitem [{\citenamefont {{Anderlini}}\ \emph {et~al.}(2007)\citenamefont
  {{Anderlini}}, \citenamefont {{Lee}}, \citenamefont {{Brown}}, \citenamefont
  {{Sebby-Strabley}}, \citenamefont {{Phillips}},\ and\ \citenamefont
  {{Porto}}}]{anderlini2007}%
  \BibitemOpen
  \bibfield  {author} {\bibinfo {author} {\bibfnamefont {M.}~\bibnamefont
  {{Anderlini}}}, \bibinfo {author} {\bibfnamefont {P.~J.}\ \bibnamefont
  {{Lee}}}, \bibinfo {author} {\bibfnamefont {B.~L.}\ \bibnamefont {{Brown}}},
  \bibinfo {author} {\bibfnamefont {J.}~\bibnamefont {{Sebby-Strabley}}},
  \bibinfo {author} {\bibfnamefont {W.~D.}\ \bibnamefont {{Phillips}}}, \ and\
  \bibinfo {author} {\bibfnamefont {J.~V.}\ \bibnamefont {{Porto}}},\ }\href
  {\doibase 10.1038/nature06011} {\bibfield  {journal} {\bibinfo  {journal}
  {\nat}\ }\textbf {\bibinfo {volume} {448}},\ \bibinfo {pages} {452} (\bibinfo
  {year} {2007})}\BibitemShut {NoStop}%
\bibitem [{\citenamefont {{F{\"o}lling}}\ \emph {et~al.}(2007)\citenamefont
  {{F{\"o}lling}}, \citenamefont {{Trotzky}}, \citenamefont {{Cheinet}},
  \citenamefont {{Feld}}, \citenamefont {{Saers}}, \citenamefont {{Widera}},
  \citenamefont {{M{\"u}ller}},\ and\ \citenamefont {{Bloch}}}]{folling2007}%
  \BibitemOpen
  \bibfield  {author} {\bibinfo {author} {\bibfnamefont {S.}~\bibnamefont
  {{F{\"o}lling}}}, \bibinfo {author} {\bibfnamefont {S.}~\bibnamefont
  {{Trotzky}}}, \bibinfo {author} {\bibfnamefont {P.}~\bibnamefont
  {{Cheinet}}}, \bibinfo {author} {\bibfnamefont {M.}~\bibnamefont {{Feld}}},
  \bibinfo {author} {\bibfnamefont {R.}~\bibnamefont {{Saers}}}, \bibinfo
  {author} {\bibfnamefont {A.}~\bibnamefont {{Widera}}}, \bibinfo {author}
  {\bibfnamefont {T.}~\bibnamefont {{M{\"u}ller}}}, \ and\ \bibinfo {author}
  {\bibfnamefont {I.}~\bibnamefont {{Bloch}}},\ }\href {\doibase
  10.1038/nature06112} {\bibfield  {journal} {\bibinfo  {journal} {\nat}\
  }\textbf {\bibinfo {volume} {448}},\ \bibinfo {pages} {1029} (\bibinfo {year}
  {2007})}\BibitemShut {NoStop}%
\bibitem [{\citenamefont {{Trotzky}}\ \emph {et~al.}(2008)\citenamefont
  {{Trotzky}}, \citenamefont {{Cheinet}}, \citenamefont {{F{\"o}lling}},
  \citenamefont {{Feld}}, \citenamefont {{Schnorrberger}}, \citenamefont
  {{Rey}}, \citenamefont {{Polkovnikov}}, \citenamefont {{Demler}},
  \citenamefont {{Lukin}},\ and\ \citenamefont {{Bloch}}}]{trotzky2008}%
  \BibitemOpen
  \bibfield  {author} {\bibinfo {author} {\bibfnamefont {S.}~\bibnamefont
  {{Trotzky}}}, \bibinfo {author} {\bibfnamefont {P.}~\bibnamefont
  {{Cheinet}}}, \bibinfo {author} {\bibfnamefont {S.}~\bibnamefont
  {{F{\"o}lling}}}, \bibinfo {author} {\bibfnamefont {M.}~\bibnamefont
  {{Feld}}}, \bibinfo {author} {\bibfnamefont {U.}~\bibnamefont
  {{Schnorrberger}}}, \bibinfo {author} {\bibfnamefont {A.~M.}\ \bibnamefont
  {{Rey}}}, \bibinfo {author} {\bibfnamefont {A.}~\bibnamefont
  {{Polkovnikov}}}, \bibinfo {author} {\bibfnamefont {E.~A.}\ \bibnamefont
  {{Demler}}}, \bibinfo {author} {\bibfnamefont {M.~D.}\ \bibnamefont
  {{Lukin}}}, \ and\ \bibinfo {author} {\bibfnamefont {I.}~\bibnamefont
  {{Bloch}}},\ }\href {\doibase 10.1126/science.1150841} {\bibfield  {journal}
  {\bibinfo  {journal} {Science}\ }\textbf {\bibinfo {volume} {319}},\ \bibinfo
  {pages} {295} (\bibinfo {year} {2008})}\BibitemShut {NoStop}%
\bibitem [{\citenamefont {{Kuklov}}\ and\ \citenamefont
  {{Svistunov}}(2003)}]{kuklov2003}%
  \BibitemOpen
  \bibfield  {author} {\bibinfo {author} {\bibfnamefont {A.~B.}\ \bibnamefont
  {{Kuklov}}}\ and\ \bibinfo {author} {\bibfnamefont {B.~V.}\ \bibnamefont
  {{Svistunov}}},\ }\href {\doibase 10.1103/PhysRevLett.90.100401} {\bibfield
  {journal} {\bibinfo  {journal} {Phys. Rev. Lett.}\ }\textbf {\bibinfo
  {volume} {90}},\ \bibinfo {eid} {100401} (\bibinfo {year}
  {2003})}\BibitemShut {NoStop}%
\bibitem [{\citenamefont {{Duan}}\ \emph {et~al.}(2003)\citenamefont {{Duan}},
  \citenamefont {{Demler}},\ and\ \citenamefont {{Lukin}}}]{duan2003}%
  \BibitemOpen
  \bibfield  {author} {\bibinfo {author} {\bibfnamefont {L.-M.}\ \bibnamefont
  {{Duan}}}, \bibinfo {author} {\bibfnamefont {E.}~\bibnamefont {{Demler}}}, \
  and\ \bibinfo {author} {\bibfnamefont {M.~D.}\ \bibnamefont {{Lukin}}},\
  }\href {\doibase 10.1103/PhysRevLett.91.090402} {\bibfield  {journal}
  {\bibinfo  {journal} {Phys. Rev. Lett.}\ }\textbf {\bibinfo {volume} {91}},\
  \bibinfo {eid} {090402} (\bibinfo {year} {2003})}\BibitemShut {NoStop}%
\bibitem [{\citenamefont {{Briegel}}\ and\ \citenamefont
  {{Raussendorf}}(2001)}]{briegel2001}%
  \BibitemOpen
  \bibfield  {author} {\bibinfo {author} {\bibfnamefont {H.~J.}\ \bibnamefont
  {{Briegel}}}\ and\ \bibinfo {author} {\bibfnamefont {R.}~\bibnamefont
  {{Raussendorf}}},\ }\href {\doibase 10.1103/PhysRevLett.86.910} {\bibfield
  {journal} {\bibinfo  {journal} {Phys. Rev. Lett.}\ }\textbf {\bibinfo
  {volume} {86}},\ \bibinfo {pages} {910} (\bibinfo {year} {2001})}\BibitemShut
  {NoStop}%
\bibitem [{\citenamefont {{Mandel}}\ \emph {et~al.}(2003)\citenamefont
  {{Mandel}}, \citenamefont {{Greiner}}, \citenamefont {{Widera}},
  \citenamefont {{Rom}}, \citenamefont {{H{\"a}nsch}},\ and\ \citenamefont
  {{Bloch}}}]{mandel2003}%
  \BibitemOpen
  \bibfield  {author} {\bibinfo {author} {\bibfnamefont {O.}~\bibnamefont
  {{Mandel}}}, \bibinfo {author} {\bibfnamefont {M.}~\bibnamefont {{Greiner}}},
  \bibinfo {author} {\bibfnamefont {A.}~\bibnamefont {{Widera}}}, \bibinfo
  {author} {\bibfnamefont {T.}~\bibnamefont {{Rom}}}, \bibinfo {author}
  {\bibfnamefont {T.~W.}\ \bibnamefont {{H{\"a}nsch}}}, \ and\ \bibinfo
  {author} {\bibfnamefont {I.}~\bibnamefont {{Bloch}}},\ }\href {\doibase
  10.1038/nature02008} {\bibfield  {journal} {\bibinfo  {journal} {\nat}\
  }\textbf {\bibinfo {volume} {425}},\ \bibinfo {pages} {937} (\bibinfo {year}
  {2003})}\BibitemShut {NoStop}%
\bibitem [{\citenamefont {{Burkard}}\ \emph {et~al.}(1999)\citenamefont
  {{Burkard}}, \citenamefont {{Loss}},\ and\ \citenamefont
  {{Divincenzo}}}]{burkard1999}%
  \BibitemOpen
  \bibfield  {author} {\bibinfo {author} {\bibfnamefont {G.}~\bibnamefont
  {{Burkard}}}, \bibinfo {author} {\bibfnamefont {D.}~\bibnamefont {{Loss}}}, \
  and\ \bibinfo {author} {\bibfnamefont {D.~P.}\ \bibnamefont {{Divincenzo}}},\
  }\href {\doibase 10.1103/PhysRevB.59.2070} {\bibfield  {journal} {\bibinfo
  {journal} {\prb}\ }\textbf {\bibinfo {volume} {59}},\ \bibinfo {pages} {2070}
  (\bibinfo {year} {1999})}\BibitemShut {NoStop}%
\bibitem [{\citenamefont {{Petta}}\ \emph {et~al.}(2005)\citenamefont
  {{Petta}}, \citenamefont {{Johnson}}, \citenamefont {{Taylor}}, \citenamefont
  {{Laird}}, \citenamefont {{Yacoby}}, \citenamefont {{Lukin}}, \citenamefont
  {{Marcus}}, \citenamefont {{Hanson}},\ and\ \citenamefont
  {{Gossard}}}]{petta2005}%
  \BibitemOpen
  \bibfield  {author} {\bibinfo {author} {\bibfnamefont {J.~R.}\ \bibnamefont
  {{Petta}}}, \bibinfo {author} {\bibfnamefont {A.~C.}\ \bibnamefont
  {{Johnson}}}, \bibinfo {author} {\bibfnamefont {J.~M.}\ \bibnamefont
  {{Taylor}}}, \bibinfo {author} {\bibfnamefont {E.~A.}\ \bibnamefont
  {{Laird}}}, \bibinfo {author} {\bibfnamefont {A.}~\bibnamefont {{Yacoby}}},
  \bibinfo {author} {\bibfnamefont {M.~D.}\ \bibnamefont {{Lukin}}}, \bibinfo
  {author} {\bibfnamefont {C.~M.}\ \bibnamefont {{Marcus}}}, \bibinfo {author}
  {\bibfnamefont {M.~P.}\ \bibnamefont {{Hanson}}}, \ and\ \bibinfo {author}
  {\bibfnamefont {A.~C.}\ \bibnamefont {{Gossard}}},\ }\href {\doibase
  10.1126/science.1116955} {\bibfield  {journal} {\bibinfo  {journal}
  {Science}\ }\textbf {\bibinfo {volume} {309}},\ \bibinfo {pages} {2180}
  (\bibinfo {year} {2005})}\BibitemShut {NoStop}%
\bibitem [{\citenamefont {{Korzh}}\ \emph {et~al.}(2015)\citenamefont
  {{Korzh}}, \citenamefont {{Lim}}, \citenamefont {{Houlmann}}, \citenamefont
  {{Gisin}}, \citenamefont {{Li}}, \citenamefont {{Nolan}}, \citenamefont
  {{Sanguinetti}}, \citenamefont {{Thew}},\ and\ \citenamefont
  {{Zbinden}}}]{korzh2015}%
  \BibitemOpen
  \bibfield  {author} {\bibinfo {author} {\bibfnamefont {B.}~\bibnamefont
  {{Korzh}}}, \bibinfo {author} {\bibfnamefont {C.~C.~W.}\ \bibnamefont
  {{Lim}}}, \bibinfo {author} {\bibfnamefont {R.}~\bibnamefont {{Houlmann}}},
  \bibinfo {author} {\bibfnamefont {N.}~\bibnamefont {{Gisin}}}, \bibinfo
  {author} {\bibfnamefont {M.~J.}\ \bibnamefont {{Li}}}, \bibinfo {author}
  {\bibfnamefont {D.}~\bibnamefont {{Nolan}}}, \bibinfo {author} {\bibfnamefont
  {B.}~\bibnamefont {{Sanguinetti}}}, \bibinfo {author} {\bibfnamefont
  {R.}~\bibnamefont {{Thew}}}, \ and\ \bibinfo {author} {\bibfnamefont
  {H.}~\bibnamefont {{Zbinden}}},\ }\href {\doibase 10.1038/nphoton.2014.327}
  {\bibfield  {journal} {\bibinfo  {journal} {Nature Photonics}\ }\textbf
  {\bibinfo {volume} {9}},\ \bibinfo {pages} {163} (\bibinfo {year}
  {2015})}\BibitemShut {NoStop}%
\bibitem [{\citenamefont {{Nielsen}}\ and\ \citenamefont
  {{Chuang}}(2010)}]{nielsen2010}%
  \BibitemOpen
  \bibfield  {author} {\bibinfo {author} {\bibfnamefont {M.~A.}\ \bibnamefont
  {{Nielsen}}}\ and\ \bibinfo {author} {\bibfnamefont {I.~L.}\ \bibnamefont
  {{Chuang}}},\ }\href@noop {} {\emph {\bibinfo {title} {Quantum Computation
  and Quantum Information, by Michael A.~Nielsen, Isaac L.~Chuang}}}\ (\bibinfo
   {publisher} {Cambridge, UK: Cambridge University Press},\ \bibinfo {year}
  {2010})\BibitemShut {NoStop}%
\bibitem [{\citenamefont {Itin}\ and\ \citenamefont
  {Katsnelson}(2015)}]{itin2014}%
  \BibitemOpen
  \bibfield  {author} {\bibinfo {author} {\bibfnamefont {A.~P.}\ \bibnamefont
  {Itin}}\ and\ \bibinfo {author} {\bibfnamefont {M.~I.}\ \bibnamefont
  {Katsnelson}},\ }\href {\doibase 10.1103/PhysRevLett.115.075301} {\bibfield
  {journal} {\bibinfo  {journal} {Phys. Rev. Lett.}\ }\textbf {\bibinfo
  {volume} {115}},\ \bibinfo {pages} {075301} (\bibinfo {year}
  {2015})}\BibitemShut {NoStop}%
\end{thebibliography}%

\end{document}